\documentstyle[aps]{revtex}
\begin{document}
\title{Large Magnetic-Field-Induced Strains in Ni-Mn-Ga Alloys due to
Redistribution of Martensite Variants}
\author{K. Ullakko}
\address{Helsinki University of Technology, Laboratory of Biomedical Engineering,\\
P.O. Box 2200, FIN-02015 HUT, Finland.}
\author{A. Sozinov}
\address{Helsinki University of Technology, Laboratory of Biomedical Engineering,\\
P.O. Box 2200, FIN-02015 HUT, Finland.}
\author{P. Yakovenko}
\address{Institute for Metal Physics, Laboratory of Thermo-Mechanical Treatment,\\
UA-252680, Kyiv 142, Ukraine. }
\maketitle

\begin{abstract}
A polycrystalline and single-crystal samples of near-stoichiometric Ni$_2$%
MnGa alloys have been investigated. It was confirmed that martensite short
crystallographic axis (c-axis) is the easy axis of magnetization. The
reversible reorientation of the easy magnetization direction of martensite
samples after 5\% compression is found. The same reversible reorientation of
the easy magnetization direction under applied magnetic field occurs in the
single-crystal samples. The field-induced reorientation of the easy
magnetization direction is accompanied with more than 4\% of sample's
dimension change. About the same value of the field-induced strains can be
obtained for polycrystalline samples after appropriate thermo-mechanical
treatment.
\end{abstract}

\section{Introduction}

Magnetic field control of the shape of ferromagnetic alloys with martensitic
structure was suggested recently \cite{c1,c2} and advantages of using those
kind of materials as actuators was substantiated \cite{c3,c4}. The
suggestion was based on the fact that crystallographic anisotropy of
martensite lattice encourages the presence of strong magnetic anisotropy.
For example, in Ni$_2$MnGa martensite lattice the easy axis of magnetization
is directed along the axis of tetragonality (c-axis) with the shortest axial
length \cite{c5}. The martensitic crystal consists of mixture of the
martensite variants possessing different c-axis orientations. The increase
in the fraction of the martensite variants with c-axis directed along the
magnetic field due to the movement of twin boundaries would appear to be
more advantageous process in comparison with the process of magnetic moment
rotation in the martensitic variants, in which the easy axis of
magnetization is non-parallel to the applied magnetic field. The change of
the martensite variants proportions would lead to macroscopic deformation of
sample, i.e. to the control of the sample shape by the applied magnetic
field.

More than 4 \% field-induced strains have been observed in a single-crystal
of Ni$_2$MnGa alloy \cite{c5,c6}. It have been shown that the initial state
of samples is important for observing such a big effect. The single
martensite variant state was obtained by applying axial compressive stress
to martensite crystal \cite{c5} or by cooling the sample through the
austenite-to-martensite phase transformation under axial compressive stress 
\cite{c6}. Magnetic field applied in transverse direction produced more than
4 \% macroscopic deformation.

The aim of this work was to confirm the coincidence of martensite c-axis
with the axis of easy magnetization in near-stoichiometric Ni$_2$MnGa alloys
and to investigate the conditions when martensite variants can be
redistributed by the applied magnetic field.

\section{Experimental}

Two alloys Ni$_{50}$Mn$_{29.5}$Ga$_{20.2}$ (N1) and Ni$_{48.7}$Mn$_{30.1}$Ga$%
_{21.3}$ (N2) were melted in induction furnace in argon atmosphere. The
composition of the alloys was measured by wave-length dispersive
spectroscopy (WDS). After homogenization at 1000 $^0$C during 3 days and
aging at 800 $^0$C 1 day alloys were air cooled to room temperature. X-ray
diffraction measurements revealed the Heusler type ordered structure (L2$_1$%
) for both alloys in austenitic state. Martensitic transformation points M$%
_s $, M$_f$, A$_s$, A$_f$, and Curie point T$_c$ were measured using low
field ac magnetic susceptibility technique (for N1 - M$_s$=71 $^0$C M$_f$=67 
$^0$C A$_s$=74 $^0$C A$_f$=78 $^0$C T$_c$=97 $^0$C, for N2 - M$_s$=29 $^0$C M%
$_f$=26 $^0$C A$_s$=32 $^0$C A$_f$=35 $^0$C T$_c$=99 $^0$C). Samples for
magnetic investigation had been cut by spark cutting machine with dimensions
4x4x4 mm. The specimens of alloy N1 had a rather fine grain structure, but
alloy N2 samples have been cut from one big grain.

All M-H curves were obtained using a vibrating sample magnetometer (VSM).
The specimen stage was mounted between the pole pieces of a 1 T
electromagnet and could be rotated for different field orientations. In all
measurements rotation of the samples was carried out at zero magnetic field.
The mechanical and thermo-mechanical treatment were carried out by
compression using tensile machine L1000R with cross head speed of 0.2
mm/min. Stains induced by magnetic field were measured by strain gauges.
Temperature was controlled with an alcohol-based circulator.

\section{Results and discussion}

Figure 1 shows the M-H curves at 25 $^0$C for alloy N1 along the different
cubic directions of the sample, marked as A, B and C, before and after 5 \%
compression. The model describing the martensite magnetization curves have
been proposed recently by Likhachev and Ullakko \cite{c7}. Discussion of our
experimental data is based on this model.

The M-H curves of the sample in three above-mentioned directions before
compression tests are shown in Fig. 1a. It is clear from the model that C
direction of the sample is close to the hard magnetization direction. For
two other sample directions A and B the M-H curves indicate the different
volume fractions of martensite variants, in which the easy axis of
magnetization is directed along the field.

The M-H curves in Fig. 1b show that the 5 \% compression of the sample along
B direction results in the sample texture, which is very similar to that for
the single variant martensite with c-axis, which is indeed the easy axis of
magnetization, along the compression direction. Along the A and C directions
M-H curves (see Fig. 1b) reveal the direction of hard magnetization.

The subsequent compression of the sample along direction A leads to the
axial texture reorientation of about 90$^0$, thus the c-axis and
consequently the easy axis of magnetization again coincides with a
compression direction (see Fig. 1c). The direction B becomes close to the
direction of hard magnetization. The change of compression direction from B
to A and vice versa results in reversible reorientation of the martensitic
texture. At the same time C direction always remains the direction of hard
magnetization. This can be connected with realization of only two
martensitic twin variants in studied sample after such a type of mechanical
treatment.

Uniaxial anisotropic constant estimated from the area between M-H curves for
easy and hard axes (Fig. 1b, 1c) was K$_u$=0.25 MJ/m3. This value is close
to that obtained by Tickle and James \cite{c5}.

Figure 2 shows the M-H curves for the single-crystal sample of alloy N2
without any preliminary treatment. The dimensions of the sample and the
procedure of measurements were the same as in previous case. The M-H curves
measured in A direction are shown in Fig. 2a. Initial part of magnetization
curve (marked as 1 in Fig. 2a) corresponds to the case where the fraction of
the martensite variant with hard axis of magnetization along A direction is
dominant (compare curves in Fig. 1 and 2). At the field in the interval
0.3-0.5 T magnetization abruptly increases and approaches the saturation
level. When decreasing the field, the magnetization keeps the saturation
level towards the lower field. Subsequent magnetization loops have no
particularities and show saturation at low field, indicating that the easy
axis of magnetization now is directed along the field (curve 2 on Fig. 2a).
When sample is rotated by 90$^0$ from A to B direction along the field, M-H
curve again shows the same behavior (see Fig. 2b). The effect is fully
reversible - 90$^0$ rotation of the sample from B to A position and vice
versa does not change the peculiarities of M-H curves. Consequently, it is
evident that under the influence of the magnetic field within the interval
of 0.3-0.5 T one can observe the easy axis reorientation from the
perpendicular to parallel to the field direction. It must be pointed out
that if the magnetic field is applied along the C direction of the sample,
no changes of easy axis direction occurs and C direction always coincides
with the hard axis of magnetization.

The above-mentioned reorientation of the easy axis is caused by the
redistribution of the martensite variants proportions under the applied
magnetic field and would lead to macroscopic deformation of the sample. The
results presented in Fig. 3 confirm this conclusion. Fig. 3a and Fig. 3b
show the changes of strain against magnetic field of alloy N2 for the
single-crystal and polycrystalline samples, respectively. For the single
crystal sample the strain gauge was attached to the preliminary magnetized
sample along A direction coincided with the easy axis of magnetization. The
polycrystalline sample was preliminary thermo-mechanically treated by 5 \%
compression cycling at the temperature higher than Af temperature with
subsequent cooling lower M$_f$ temperature. X-ray diffraction measurements
of the cubic structure, made before thermo-mechanical treatment, revealed
that preferable orientation along columnar grains was the 
\mbox{$<$}
001%
\mbox{$>$}
. The strain gauge was attached to the polycrystalline sample along the
compression direction, which was parallel to the columnar grains.

Curves 1 and 2 in Fig. 3 show the changes of the strain vs. magnetic field
applied across the strain gauge axis. Then the sample was turned by 90$^0$
at zero field (B single crystal sample direction is parallel to the magnetic
field), and subsequently the strain was measured as a function of the
magnetic field (curves 3 and 4). Increasing the magnetic field after every 90%
$^0$ rotation of the sample results in more than 4 \% strain. The changes of
single crystal sample dimensions due to rotation between A and B directions
are reversible, but applying the magnetic field along direction C causes no
dimension changes. Results shown in Fig. 2 and Fig. 3a are in a good
accordance to each other. The difference between curves 1 in the Figs. 3a
and 3b can be connected with that the polycrystalline sample has not been
magnetized before measurements. The rather big sample shape changes give us
the possibility to confirm strain gauge data by micrometer measurements. The
observed magnetic-field induced reversible strains are the largest to date
obtained for polycrystalline Ni$_2$MnGa alloys.

\section{Conclusions}

1. Compression of Ni$_2$MnGa alloy samples by approximately 5 \%, at a
temperature lover that the M$_f$ one results in formation of close to one
variant martensite structure in which c-axis coincides with compression
direction. It is confirmed that martensite c-axis is the axis of easy
magnetization.

2. In single crystal Ni$_2$MnGa alloy sample the reversible redistribution
of twin variants by applying the magnetic field can be obtained without any
preliminary treatment. The martensite structure reorientation occurs only
between two sample directions, the third one remains the axis of hard
magnetization.

3. The rearrangement of the martensite variants under influence of magnetic
field is accompanied with reversible changes of sample dimensions exceeding
4 \%. The same effect can be achieved even on polycrystalline samples after
appropriate thermo-mechanical treatment.

{\bf Acknowledgments}

Authors acknowledge the Physical and Materials Science Departments of the
Helsinki Technological University supported this work.


\begin{references}
\bibitem{c1}  K. Ullakko, Proc. Third International Conf. on Intelligent
Materials 1996, ed. P. F. Gobin and J. Tatibouet, Lyon, France, SPIE 2779
(1996), p. 505.

\bibitem{c2}  K. Ullakko, P.T. Jakovenko, V.G. Gavriljuk, Proc. Conf. on
Smart Structures and Materials, ed. V.V. Varadan and J. Chandra, San Diego,
USA, SPIE 2715 (1996), p. 42.

\bibitem{c3}  R.D. James, M. Wuttig, Proc. Conf. on Smart Structures and
Materials, ed. V.V. Varadan and J. Chandra, San Diego, USA, SPIE 2715
(1996), p. 420.

\bibitem{c4}  K. Ullakko, J. Mater. Eng. Perform. 5 (1996), p. 405.

\bibitem{c5}  R. Tickle, R.D. James, J. Magn. Magn. Mat. 195 (1999), p. 629.

\bibitem{c6}  R.D. James, R. Tickle, M. Wuttig, Int. Conf. on Martensitic
Transformations ICOMAT-98, San Carlos de Bariloche, Argentina, Poster
Presentation and Conference Abstracts (1998), p. 1.14

\bibitem{c7}  A.A. Likhachev, K. Ullakko, EPJdirect B2 (1999), p. 1.; Eur.
Phys. J. B14 (2000) p. 263.
\end{references}
\end{document}